\documentclass[prl,aps,showpacs,superscriptaddress,tightenlines,twocolumn]{revtex4}

\usepackage{epsfig}
\usepackage[latin1]{inputenc}
\usepackage{float,amsmath,slashed}
\usepackage{graphicx}

\newcommand{\beq}{\begin{equation}}
\newcommand{\eeq}{\end{equation}}
\newcommand{\bqa}{\begin{eqnarray}}
\newcommand{\eqa}{\end{eqnarray}}

\def\lsim{\mathrel{\rlap{\lower4pt\hbox{$\sim$}}
    \raise1pt\hbox{$<$}}}                
\def\gsim{\mathrel{\rlap{\lower4pt\hbox{$\sim$}}
    \raise1pt\hbox{$>$}}}                

\begin{document}

\title{Jet broadening in unstable non-Abelian plasmas}

\author{Adrian Dumitru}
\affiliation{Institut f\"ur Theoretische Physik \\
  Johann Wolfgang Goethe - Universit\"at Frankfurt \\
  Max-von-Laue-Stra\ss{}e~1,
  D-60438 Frankfurt am Main, Germany\vspace*{2mm}}
\author{Yasushi Nara}
\affiliation{Akita International University 193-2 Okutsubakidai \\
  Yuwa-Tsubakigawa, Akita-City, Akita 010-1211, Japan \\
  \vspace*{2mm}}
\author{Bj\"orn Schenke}
\author{Michael Strickland}
\affiliation{Institut f\"ur Theoretische Physik \\
  Johann Wolfgang Goethe - Universit\"at Frankfurt \\
  Max-von-Laue-Stra\ss{}e~1,
  D-60438 Frankfurt am Main, Germany\vspace*{2mm}}

\begin{abstract}
We perform numerical simulations of the SU(2) Boltzmann-Vlasov 
equation including both hard elastic particle collisions and soft 
interactions mediated by classical Yang-Mills fields.  Using this 
technique we calculate the momentum-space broadening of high-energy 
jets in real-time for both locally isotropic and anisotropic plasmas. 
In both cases we introduce a separation scale which separates hard and 
soft interactions and demonstrate that our results for jet broadening 
are independent of the precise separation scale chosen.  For an 
isotropic plasma this allows us to calculate the jet transport 
coefficient $\hat{q}$ including hard and soft non-equilibrium 
dynamics. For an anisotropic plasma the jet transport coefficient 
becomes a tensor with $\hat{q}_L \neq \hat{q}_\perp$. We find that for 
weakly-coupled anisotropic plasmas the fields develop unstable modes, 
forming configurations where $B_\perp>E_\perp$ and $E_z>B_z$ which
lead to $\hat{q}_L > \hat{q}_\perp$.
We study whether the effect is strong enough to explain the experimental 
observation that high-energy jets traversing the 
plasma perpendicular to the beam axis experience much stronger 
broadening in rapidity, $\Delta \eta$, than in azimuth, $\Delta \phi$.
\end{abstract}
\maketitle
\newpage

\section{Introduction}
High transverse momentum jets produced in heavy-ion collisions 
represent a valuable tool for studies of the properties of the hot 
parton plasma produced in the central rapidity 
region~\cite{Jacobs:2005pk}.  This is due to the fact that jets couple 
to the plasma causing the jet to broaden in momentum space and to lose 
energy.  The magnitude of momentum-space broadening and energy loss 
experienced by a parton depends on whether or not one assumes the 
matter to be hadronic or partonic in nature. Hence, it is one of the 
primary observables to ascertain experimentally whether or not the plasma has 
been produced. At very high energies it is expected 
that hard bremsstrahlung processes dominate the light quark or gluon 
energy loss~\cite{Baier:1996sk}; however, at intermediate energies the 
inclusion of both collisional and radiative processes is necessary in 
order to make phenomenological predictions. Here we present first 
results from real-time solution of the SU(2) Boltzmann-Vlasov equation 
for locally isotropic and anisotropic plasmas which include both hard 
(scattering) and soft (classical field) processes.

We first demonstrate that in a locally isotropic plasma, one can
obtain a cutoff-independent transport coefficient $\hat{q}$,
which measures the square of the transverse momentum transfer per mean
free path. This measurement lays the groundwork for determining other
jet transport properties like the energy-loss spectrum from real-time
simulations. We then perform a similar measurement in a locally
anisotropic plasma and demonstrate that for a non-Abelian plasma
unstable modes can cause asymmetric broadening of jets. This may be
relevant for recent measurements of di-hadron correlations which
provide evidence for an {\em asymmetric} broadening of jet profiles in
the plane of pseudorapidity ($\eta$) and azimuthal angle ($\phi$),
with $\Delta \eta > \Delta\phi$, which has been called ``the ridge''
by experimentalists~\cite{Jacobs:2005pk,Putschke:2007mi}.  Here we
will show that this asymmetry could partly be caused by unstable
plasma modes which are induced by the longitudinal expansion of the
plasma.

At the earliest times after an ultrarelativistic heavy-ion collision, 
before thermalization and hydrodynamic expansion, the 
plasma undergoes rapid longitudinal expansion.  This expansion can 
lead to an oblate anisotropic  ($\langle p_z^2\rangle \ll \langle 
p_\perp^2\rangle$) momentum distribution in the local rest 
frame~\cite{bottomUp,Romatschke:2003ms}. It has been shown that
instabilities develop~\cite{Romatschke:2003ms} in such anisotropic
plasmas which lead to the formation of 
long-wavelength chromo-magnetic and chromo-electric fields.  These 
fields can then affect the propagation of hard jets and their 
induced hard radiation field.  In the Abelian case soft transverse 
magnetic fields, $B_\perp$, dominate all other field components and 
therefore in this case one expects the longitudinal pressure induced 
by unstables modes to be larger than the transverse 
pressure~\cite{Romatschke:2006bb,Majumder:2006wi}.  
This pressure asymmetry causes asymmetric broadening of jets with 
larger broadening along the longitudinal or $\eta$ direction.  

The situation is more complicated in non-Abelian plasmas since then, in 
addition to generating large coherent $B_\perp$ and $E_\perp$ domains, 
one also generates large-amplitude longitudinal fields $B_z$ and 
$E_z$.  It is therefore not obvious a priori that the pressure 
generated by other field components will result in an asymmetric 
broadening of the jet. Through our numerical simulations we find that 
for oblate parton momentum distributions, that at different times 
either $E_z>B_z$ or $B_\perp >E_\perp$ with the net effect being 
a factor of 1.5 stronger longitudinal than transverse broadening.

\section{Boltzmann-Vlasov equation for non-Abelian gauge theories}
We solve the classical transport equation for hard gluons with SU(2)
color charge $q=q^a t^a$~\cite{Heinz:1983nx}, with the color generators $t^a$, including hard binary
collisions
\begin{align}
    p^\mu \left[\partial_\mu + g q^a F_{\mu\nu}^a \partial^\nu_{p} 
    + g f^{abc} A_\mu^b(x) q^c \partial_{q^a} \right]f={\cal C}\,,
\end{align}
where $f=f(x,p,q)$ denotes the single-particle phase space distribution.
It is coupled self-consistently to the Yang-Mills equation for the
soft gluon fields,
\begin{equation}
 D_\mu F^{\mu\nu} = j^\nu 
 = g \int \frac{d^3p}{(2\pi)^3}\,dq\,q\,v^\nu\,f(x,p,q)\,,
\end{equation}
with $v^\mu=(1,\mathbf{p}/p)$.  When the phase-space density is
parametrically small, $f={\cal O}(1)$, the collision term is given by
\begin{align}
\label{cterm1}
{\cal C} &= \frac{1}{4E_1} \int_{\mathbf{p_2}}\,
\int_{\mathbf{p'_1}} \int_{\mathbf{p'_2}}(2\pi)^4\delta^{(4)}
(p'_1+p'_2-p_1-p_2)\notag \\ 
&~~~~~~~~\times\left(f'_1 f'_2 |{\cal M}_{1'2'\to 12} |^2 -  f_1 f_2
|{\cal M}_{12\to 1'2'}|^2 \right) \,,
\end{align}
with $\int_{\mathbf{p_i}}=\int \frac{d^3p_i}{(2\pi)^3 2E_i}$. The
matrix element ${\cal M}$ includes all $gg\rightarrow gg$ tree-level
diagrams and color factors as appropriate for the SU(2) gauge group.

We employ the test particle method to replace the
continuous distribution $f(x,p,q)$ by a large number of test
particles, which
leads to Wong's equations~\cite{Wong:1970fu}
\begin{align}
    \dot{\mathbf{x}}_i(t)&=\mathbf{v}_i(t)\,, \label{wong1}\\
    \dot{\mathbf{p}}_i(t)&=g
    q^a_i(t)\left(\mathbf{E}^a(t)+\mathbf{v}_i(t)\times\mathbf{B}^a(t)\right)
    \,,\label{wong2}\\ 
    \dot{q}_i(t)&=-igv_i^\mu(t)[A_\mu(t),q_i(t)]\,,\label{wong3}
\end{align}
for the $i$-th test particle, whose coordinates are $\mathbf{x}_i(t)$,
$\mathbf{p}_i(t)$, and $q^a_i(t)$. The time evolution of the
Yang-Mills field is determined by the standard Hamiltonian method
\cite{Ambjorn:1990pu} in $A^0=0$ gauge. Our lattices have periodic
boundary conditions and a lattice spacing $a$, whose physical value is
fixed by the length of the lattice $L=aN_s$.  Dimensionless lattice
variables scale such that when the number $N_s$ of lattice sites is
changed, $L$ remains fixed; see~\cite{Dumitru:2005gp,Dumitru:2006pz}
for details. 

The theory without collisions as given by
equations~(\ref{wong1}-\ref{wong3}) coupled to the lattice Yang-Mills
equations was first solved in ref.~\cite{HuBM} to study Chern-Simons
number diffusion in non-Abelian gauge theories at finite temperature.
It was applied later also to the problem of gauge-field instabilities
in anisotropic SU(2) plasmas~\cite{Dumitru:2005gp,Dumitru:2006pz}. Our
numerical implementation is based on the improved formulation detailed
in ref.~\cite{Dumitru:2006pz} where the non-Abelian currents generated
by the hard particle modes on the lattice sites are ``smeared'' in
time. This technique makes simulations in three dimensions on large
lattices possible in practice.

In this paper we go beyond those earlier simulations by accounting
also for hard (short-distance) collisions among particles.
The collision term is incorporated using the stochastic method
\cite{Danielewicz:1991dh}. Scattering processes are determined by
sampling possible transitions according to the collision
rate in a lattice cell:
\begin{equation}
\label{p22test}
\frac{dP_{2\to2}}{dt} = 
  \tilde{v}_{\text{rel}} \frac{\sigma_{2\to2}}{a^3 N_{\text{test}}}\,, 
\end{equation}
with $\tilde{v}_{\text{rel}}=s/(2E_1E_2)$, where $s$ is the invariant
mass of a gluon pair. The total cross section is given by
\begin{equation}
    \label{totcross}
    \sigma_{2\to2}=\int_{k^{*2}}^{{s}/{2}}\frac{d\sigma}{dq^2}dq^2\,.
\end{equation}
The momentum transfer is determined in the center of mass frame of the
two colliding particles from the probability distribution
\begin{equation}
 {\cal P}(q^2)=\frac{1}{\sigma_{2\to2}}\, \frac{d\sigma}{dq^2}\,.
\end{equation}
In Eq.~(\ref{totcross}) we have introduced an infrared cutoff $k^*$ for
point-like binary collisions. To avoid double-counting, this cutoff
should be on the order of the hardest field mode that can be
represented on the given lattice, $k^*\simeq\pi/a$. Momentum
transfers below $k^*$ are mediated by the classical Yang-Mills field;
a soft scattering corresponds to deflection of a particle in the
field of the other(s). Note, that we use the color averaged expression for the collision term.
The color charge of a particle is hence not affected by a hard collision.

Physically, the separation scale $k^*$ should be sufficiently small so
that the soft field modes below $k^*$ are highly
occupied~\cite{Ambjorn:1990pu}. On the other hand, $k^*$ should be
sufficiently large to ensure that hard modes can be represented by
particles and that collisions are described correctly
by Eq.\,(\ref{cterm1}). Furthermore, unstable modes arise in anisotropic
plasmas (see below), all of which should be located below $k^*$.
Since $g\sim1$, in practice, we choose $k^*=\sqrt{3}\pi/a$ to be on the
order of the temperature for isotropic systems, and on the order of
the hard transverse momentum scale for anisotropic
plasmas. Independently, one should have $m_\infty L\gg1$ and $m_\infty
a\ll 1$: the first condition ensures that the relevant soft modes
actually fit on the lattice while the latter corresponds to the
continuum limit. Here, $m_\infty$ denotes the
soft scale and is given by
\begin{equation} \label{minf}
m_\infty^2 = g^2N_c\int\frac{d^3p}{(2\pi)^3}\frac{f(\mathbf{p})}
 {|\mathbf{p}|} \sim g^2N_c\frac{n_g}{p_{\text{h}}}\,,
\end{equation}
where $N_c=2$ is the number of colors and $n_g$ denotes the number
density of hard gluons, summed over two helicities and $N_c^2-1$
colors. Also, $p_{\text{h}}\approx 3T$ is the typical momentum of
a hard particle from the medium.

As we have argued above, we shall choose the inverse lattice spacing
to be on the order of the temperature of the medium. Thus,
with~(\ref{minf}) the continuum condition $m_\infty a \ll 1$ roughly
translates into
\begin{equation} \label{eq:continuum}
g^2 N_c \, \frac{n_g}{T^3} \ll 1~.
\end{equation}
In order to satisfy this relation with $g\sim1$, in our numerical
simulations below we shall assume an extremely hot medium, $T^3\gg
n_g$. However, this should be viewed simply as a numerical procedure
which ensures that the simulations are carried out near the continuum
(or weak-coupling) limit. We shall verify below that transverse
momentum broadening of a high-energy jet passing through a thermal
medium is independent of $T$ if the density and the ratio of jet
momentum to temperature is fixed; compare to Eq.\,(\ref{<kt2>LL})
below. One may therefore obtain a useful ``weak-coupling'' estimate of
$\langle p_\perp^2\rangle$ (resp.\ for the related transport
coefficient $\hat{q}$) by extrapolating our measurements down to
realistic temperatures.


\section{Jet broadening in an isotropic plasma}
We first consider a heat-bath of particles with a density of
$n_g=10/\text{fm}^3$ and an average particle momentum of
$3T=12$~GeV. The rather extreme ``temperature'' is chosen to satisfy
the above conditions on $N_s=32\cdots128$ lattices, assuming
$L=15$~fm. For a given lattice (resp.\ $k^*$) we take the initial
energy density of the thermalized fields to be 
\begin{equation}
\int \frac{d^3k}{(2\pi)^3} \,k \hat{f}_{\rm Bose}(k) \, \Theta(k^*-k)\,,
\end{equation}
where $\hat{f}_{\rm Bose}(k)=n_g/(2T^3 \zeta(3))/(e^{k/T}-1)$ is a
Bose distribution normalized to the assumed particle density $n_g$,
and $\zeta$ is the Riemann zeta function.  This is equivalent to the
energy density of Bose-distributed particles with momenta below the
separation momentum $k^*$. The initial spectrum is fixed to Coulomb
gauge and $A_i\sim 1/k$ (in continuum notation); also, for simplicity
we set $E_i=0$ at the initial time but electric fields build up
quickly within just a few time steps.

\begin{figure}[t]
  \begin{center}
    \includegraphics[width=7cm]{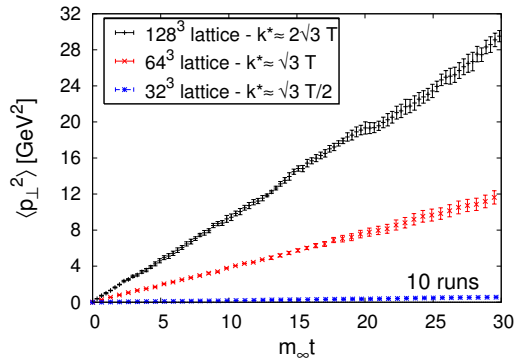}
    \caption{(Color online) Momentum diffusion caused by
    particle-field interactions only. Additional high-momentum modes
    on larger lattices cause stronger momentum broadening. $T=4$ GeV,
    $g=2$, $N_c=2$, $n_g=10/\text{fm}^3$, $m_\infty=1.4/$fm.}
    \label{fig:ptnocoll}
  \end{center}
\end{figure}

\begin{figure}[t]
  \begin{center}
    \includegraphics[width=7cm]{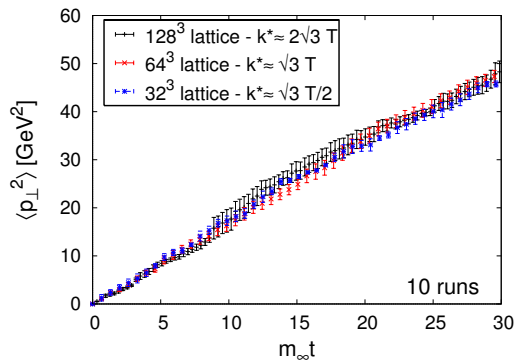}
    \caption{(Color online) Momentum diffusion by both particle-field
    and direct particle-particle interactions. The result is
    independent of the separation scale $k^*$. Same parameters as in
    Fig.~\ref{fig:ptnocoll}.}
    \label{fig:ptcoll}
  \end{center}
\end{figure}

We then measure the momentum broadening $\langle p_\perp^2\rangle(t)$
of high-energy test particles ($p/3T\approx 5$) passing through this
medium. Fig.~\ref{fig:ptnocoll} shows that in the collisionless case,
${\cal C}=0$, the broadening is stronger on larger lattices, which
accommodate harder field modes. However, Fig.~\ref{fig:ptcoll}
demonstrates that collisions with momentum exchange larger than
$k^*(a)$ compensate for this growth and lead to approximately
lattice-spacing independent results even when $k^*$ varies by a
factor of four.

Figs.~\ref{fig:ptnocoll} and \ref{fig:ptcoll} show that the relative
contributions to $\langle p_\perp^2\rangle$ from soft and hard
exchanges can depend significantly on $k^*$, even for $p/k^*={\cal
  O}(10)$. It is clear, therefore, that transport coefficients
obtained in the leading logarithmic (LL) approximation from the pure
Boltzmann approach (without soft fields) will be rather sensitive to
the infrared cutoff $k^*$. Fitting the difference of
Fig.~\ref{fig:ptcoll} and Fig.~\ref{fig:ptnocoll} (i.e., the hard
contribution) to the LL formula
\begin{equation}    \label{<kt2>LL}
  \frac{d\langle p_\perp^2\rangle}{dt}  = \frac{C_A}{C_F}
  \frac{g^4}{8\pi}  n_g \log\left(
  C^2\frac{p^2}{k^{* 2}} \right) \,,
\end{equation}
gives $C\simeq 0.43$, $0.41$, $0.31$ for $k^*/T=2\sqrt{3}$,
$\sqrt{3}$, $0.5\sqrt{3}$, respectively.
For the full calculation $C\simeq 0.61 \, k^*/(\sqrt{3}T)$.

A related and frequently used transport coefficient is $\hat
q$~\cite{Baier:1996sk}. It is the typical momentum transfer (squared)
per collision divided by the mean-free path, which is nothing but
$\langle p_\perp^2\rangle(t)/t$. From Fig.~\ref{fig:ptcoll}, $\hat
q\simeq 2.2$~GeV$^2$/fm for $N_c=2$, $n_g=10/$fm$^3$ and
$p/(3T)\approx5$.  Our cut-off independent value for $\hat q$ is in
the range extracted from phenomenological analyses of jet-quenching
data from RHIC~\cite{RHICqhat}.

\begin{figure}[t]
  \begin{center}
    \includegraphics[width=7cm]{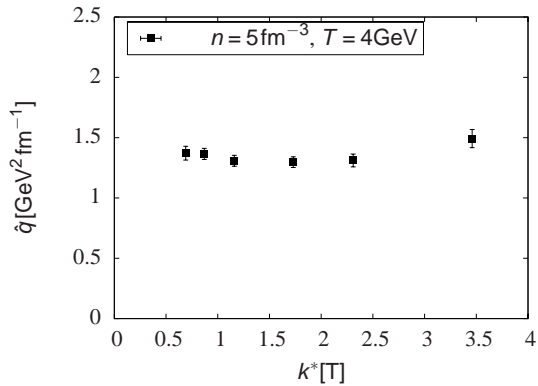}
    \caption{$\hat q$ as a function of $k^*$ at fixed $n_g=5$/fm$^3$ and $p/(3T)=16$.}
    \label{fig:qhat_k}
  \end{center}
\end{figure}

In Fig.\ref{fig:qhat_k} we show that $\hat{q}$ is indeed largely
independent of the separation scale $k^*$.  In these simulations, test
particles were explicitly bunched into colorless jets, such that
radiative energy loss does not contribute. This explains why one does
not need bremsstrahlung processes in the collision term to obtain a
cutoff independent result. Note that the magnitude of
$\hat{q}\simeq1.3$~GeV$^2/$fm at $n_g=5$/fm$^3$ and $p/(3T)=16$ is
smaller than the extrapolation obtained from
Eq.~(\ref{<kt2>LL}) with the constant under the logarithm fixed
from the previous runs at $n_g=10$/fm$^3$ and $p/(3T)=5$; hence, $C$
effectively depends on the density and the jet momentum.

\begin{figure}[t]
  \begin{center}
    \includegraphics[width=7.5cm]{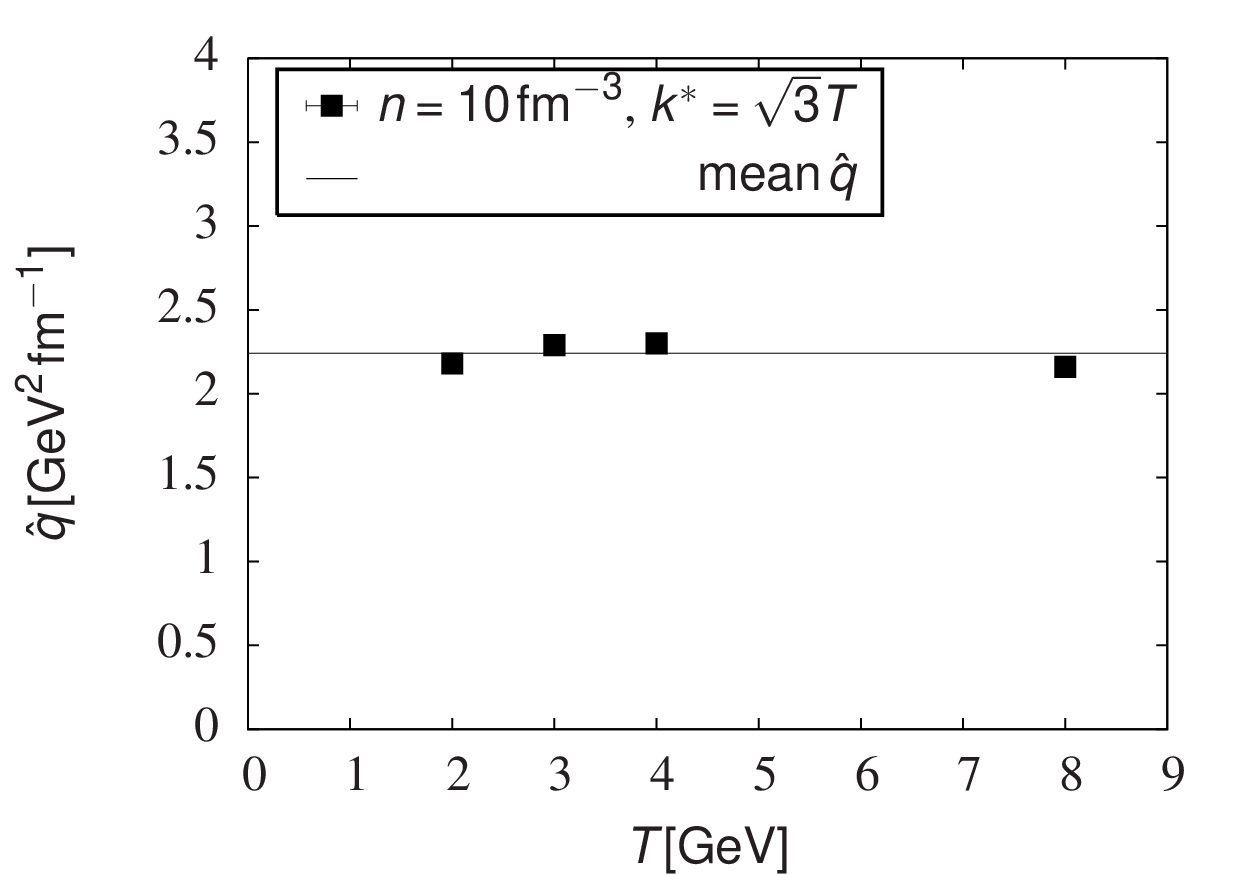}
    \caption{$\hat q$ as a function of $T$ at fixed
    $n_g=10$/fm$^3$ and $p/(3T)=5$.}
    \label{fig:qhat_T}
  \end{center}
\end{figure}

In Fig.~\ref{fig:qhat_T} we verify that $\hat q$ does not depend on
the temperature $T$ so long as the particle density $n_g$ and the
ratio of jet momentum to temperature, $p/T$, is fixed. Thus, our
measurement of $\hat q\simeq 2.2$~GeV$^2$/fm may be considered as a
weak-coupling extrapolation to realistic temperatures of
$T\approx300$~MeV and jet momenta of about 4.5~GeV.


\section{Jet broadening in an unstable plasma}
In heavy-ion collisions, locally anisotropic momentum distributions
can emerge due to the longitudinal expansion. Such anisotropies
generically give rise to instabilities~\cite{Romatschke:2003ms};
see~\cite{Dumitru:2005gp,Dumitru:2006pz} for simulations of unstable
non-Abelian plasmas within the present ``Wong-Yang-Mills'' approach.
Here, we investigate their effect on the momentum broadening of jets,
including the effect of collisions.  The initial momentum distribution
for the hard plasma gluons is taken to be
\begin{align}
\label{anisof}
f(\mathbf{p})=n_g \left(\frac{2\pi}{p_{\text{h}}}\right)^2 \delta(p_z)
  \exp(-p_\perp/p_{\text{h}})\,, 
\end{align}
with $p_\perp=\sqrt{p_x^2+p_y^2}$. This represents a quasi-thermal
distribution in two dimensions with average momentum $=2\,p_{\text{h}}$.
We initialize small-amplitude fields sampled from a
Gaussian distribution and set $k^*\approx p_{\text{h}}$, for the
reasons alluded to above. The band of unstable modes is located below
$k^*$.

We find that binary collisions among hard particles reduce the growth
rate of unstable field modes, in agreement with
expectations~\cite{Schenke:2006xu}. However, for
$p_{\text{h}}=16$~GeV, $L=5$~fm, $n_g=10/\text{fm}^3$, $g=2$,
$m_\infty\simeq1$/fm and $k^*\approx 1.7 p_{\text{h}}$, the reduction of
the growth rate is only approximately $5\%$, increasing to about $15\%$ when
$k^*\approx 0.9 p_{\text{h}}$. This is due to fewer available field
modes and more randomizing collisions.

Next, we add additional high momentum particles with $p_x=12\,
p_{\text{h}}$ and $p_x=6\, p_{\text{h}}$, respectively, to investigate
the broadening in the $y$ and $z$ directions via the variances
\begin{align}
	\hat{q}_\perp(p_x):=\frac{d}{dt}\langle(\Delta
        p_\perp)^2\rangle\,, 
      ~~~\hat{q}_L(p_x):=\frac{d}{dt}\langle(\Delta p_z)^2\rangle\,.
\end{align}
The quantity $\sqrt{\hat{q}_L/\hat{q}_\perp}$ can be roughly associated with the
ratio of jet correlation widths in azimuth and rapidity:
$\sqrt{{\hat{q}_L}/{\hat{q}_\perp}} \approx {\langle\Delta\eta\rangle} /
{\langle\Delta\phi\rangle}$.


\begin{figure}[htb]
  \begin{center}
    \includegraphics[width=7cm]{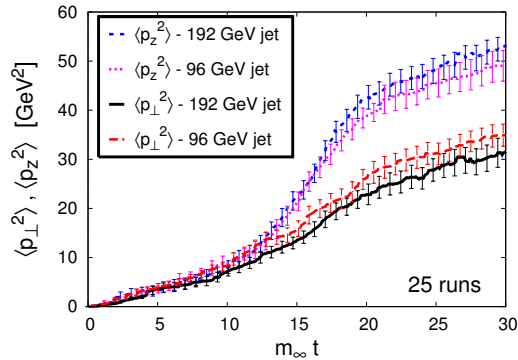}
    \caption{(Color online) Momentum broadening of a jet in the
      directions transverse to its initial momentum. $p_z$ is directed
      along the beam axis, $p_\perp$ is transverse to the
      beam. Anisotropic plasma, $64^3$ lattice.}
    \label{fig:pxyjet192}
  \end{center}
\end{figure}
Fig.~\ref{fig:pxyjet192} shows the time evolution of $\langle
p_\perp^2 \rangle$ and of $\langle p_z^2 \rangle$. The strong growth
of the soft fields sets in at about $t\simeq 10 \,m_\infty^{-1}$ and
saturates around $t\simeq 25\, m_\infty^{-1}$ due to the finite lattice
spacing (also see \cite{Dumitru:2006pz}). Outside the above time
interval the ratio $\hat{q}_L/\hat{q}_\perp\approx1$.
During the period of instability, however,
\begin{align}
	\frac{\hat{q}_L}{\hat{q}_\perp} \approx 2.3\,,
\end{align}
for both jet energies shown in Fig.~\ref{fig:pxyjet192}.  We find
approximately the same ratio for denser plasmas ($n_g=20/\text{fm}^3$
and $n_g=40/\text{fm}^3$). Reducing the number of lattice sites and
scaling $p_{\text{h}}$ down to 8~GeV gives
$\hat{q}_L/\hat{q}_\perp\approx2.1$. However, these latter runs are
rather far from the continuum limit and lattice artifacts are
significant~\cite{Dumitru:2006pz}.


The explanation for the larger broadening along the beam axis is as
follows. In the Abelian case the instability generates predominantly
transverse magnetic fields which deflect the particles in
the $z$-direction~\cite{Majumder:2006wi}.

\begin{figure}[htb]
  \begin{center}
    \includegraphics[width=7cm]{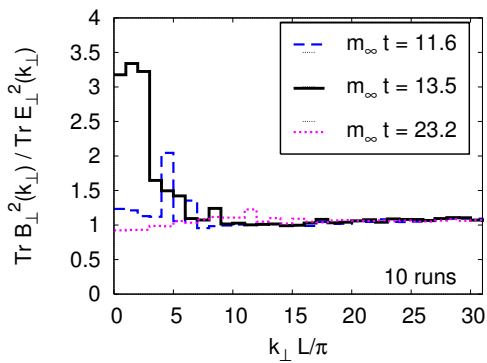}
    \caption{(Color online) Ratios of Fourier transforms of the field energy
      densities (integrated over $k_z$) at various times; the gauge
      potentials were transformed to Coulomb gauge.}
    \label{fig:FTs}
  \end{center}
\end{figure}
\begin{figure}[htb]
  \begin{center}
    \includegraphics[width=7cm]{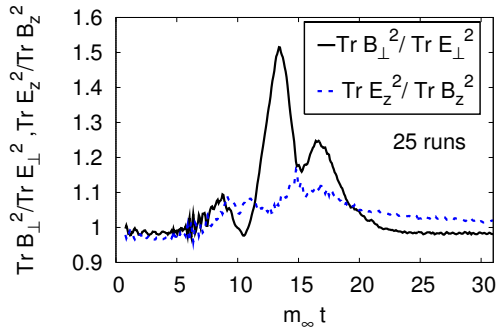}
    \caption{(Color online) Ratios of field energy densities.}
    \label{fig:ratios}
  \end{center}
\end{figure}

In the non-Abelian case, however, on three-dimensional lattices
transverse magnetic fields are much less dominant (see, e.g.\ Fig.~5
in~\cite{Dumitru:2006pz}) although they do form larger coherent
domains in the transverse plane at intermediate times than $E_\perp$,
Fig.~\ref{fig:FTs}. Longitudinal fields and locally non-zero
Chern-Simons number $\sim {\rm tr}~\mathbf{E} \cdot \mathbf{B}$
emerge, also. Nevertheless, Fig.~\ref{fig:ratios} shows that
$E_z>B_z$, aside from $B_\perp>E_\perp$.  Hence, the field
configurations are such that particles are deflected preferentially in
the longitudinal $z$-direction (to restore isotropy).

A third contribution to $p_z$ broadening in an expanding plasma, not
considered explicitly here, is due to a longitudinal collective flow
field which ``blows'' the jet fragments to the
side~\cite{Armesto:2004vz}. This mechanism is also available for
collision dominated plasmas with (nearly) isotropic momentum
distribution. However, rather strong flow gradients seem to be
required to reproduce the observed broadening of midrapidity jets (the
flow velocity has to vary substantially {\em within} the narrow jet
cone). In contrast, color fields will naturally deflect particles with
lower momentum by larger angles ($\Delta p\sim E,B$): the jet profile
broadens even if the induced radiation is exactly collinear. It is
therefore important to determine, experimentally, whether the
asymmetric broadening is related to the {\em macroscopic} collective
flow or to an anisotropy of the plasma in the local rest frame.
More detailed simulations should account, also, for the fact that
small-$x$ gluons are already correlated over large rapidity intervals
at the initial time~\cite{DGMV}.

\section*{Acknowledgments}
B.S.\ thanks O.\ Fochler, P.\ Romatschke and Z.\ Xu for many useful
discussions. The simulations were performed at the Center for
Scientific Computing of Goethe University.
B.S. also thanks the Institute for Nuclear Theory at the University of Washington for its hospitality.
M.S.\ and B.S.\ are supported by DFG Grant GR 1536/6-1.


\begin{thebibliography}{99}

\bibitem{Jacobs:2005pk} 
P.~Jacobs, Eur.\ Phys.\ J.\ \textbf{C43}, 467 (2005).

\bibitem{Baier:1996sk}
R.~Baier, Y.~L.~Dokshitzer, A.~H.~Mueller, S.~Peigne and D.~Schiff,
  Nucl.\ Phys.\  B {\bf 484}, 265 (1997).

\bibitem{Putschke:2007mi}
  F.~Wang  [STAR Collaboration],
  J.\ Phys.\ G {\bf 30}, S1299 (2004);
  J.~Adams {\it et al.}  [STAR Collaboration],
  Phys.\ Rev.\ Lett.\  {\bf 95}, 152301 (2005),
  Phys.\ Rev.\  C {\bf 73}, 064907 (2006);
  J.~Putschke,
  J.\ Phys.\ G {\bf 34}, S679 (2007).

\bibitem{bottomUp}
R.~Baier, A.~H.~Mueller, D.~Schiff and D.~T.~Son,
Phys.\ Lett.\  B {\bf 502}, 51 (2001).

\bibitem{Romatschke:2003ms}
P.~Romatschke and M.~Strickland, Phys.\ Rev.\ \textbf{D68}, 036004 (2003);
P.~Arnold, J.~Lenaghan, and G.~D.~Moore, JHEP \textbf{08}, 002 (2003);
S.~Mrowczynski,
  Acta Phys.\ Polon.\  B {\bf 37}, 427 (2006).

\bibitem{Romatschke:2006bb}
P.~Romatschke, Phys.\ Rev.\ \textbf{C75}, 014901 (2007);
P.~Romatschke and M.~Strickland,
  Phys.\ Rev.\  D {\bf 71}, 125008 (2005).

\bibitem{Majumder:2006wi}
A.~Majumder, B.~M\"uller, and S.~A.~Bass, 
  Phys.\ Rev.\ Lett.\ \textbf{99}, 042301 (2007).
\bibitem{Heinz:1983nx}
U.~W.~Heinz,
 Phys.\ Rev.\ Lett.\ \textbf{51}, 351 (1983); 
 Ann.\ Phys.\ \textbf{161}, 48 (1985);
 {\em ibid.} \textbf{168}, 148 (1986);
H.-T.~Elze and U.~W.~Heinz, Phys.\ Rept.\ \textbf{183}, 81 (1989);
J.~P.~Blaizot and E.~Iancu, Phys.\ Rev.\ Lett.\ \textbf{70}, 3376 (1993);
  Nucl.\ Phys.\ \textbf{B417}, 608 (1994);
P.~F.~Kelly, Q.~Liu, C.~Lucchesi, and C.~Manuel,
  Phys.\ Rev.\ Lett.\ \textbf{72}, 3461 (1994);
  Phys.\ Rev.\ \textbf{D50}, 4209 (1994);
D.~F.~Litim and C.~Manuel,
  Phys.\ Rev.\ Lett.\  {\bf 82}, 4981 (1999);
P.~Arnold, G.~D.~Moore and L.~G.~Yaffe,
  JHEP {\bf 0301}, 030 (2003).

\bibitem{Wong:1970fu}
S.~K.~Wong, Nuovo Cim.\ \textbf{A65}, 689 (1970).

\bibitem{Ambjorn:1990pu}
J.~Ambjorn, T.~Askgaard, H.~Porter, and M.~E.~Shaposhnikov,
  Nucl.\ Phys.\ \textbf{B353}, 346 (1991);
A.~Krasnitz and R.~Venugopalan,
  Nucl.\ Phys.\  B {\bf 557}, 237 (1999).

\bibitem{Dumitru:2005gp}
A.~Dumitru and Y.~Nara, Phys.\ Lett.\ \textbf{B621}, 89 (2005).

\bibitem{Dumitru:2006pz}
A.~Dumitru, Y.~Nara, and M.~Strickland,
  Phys.\ Rev.\ \textbf{D75}, 025016 (2007).

\bibitem{HuBM}
C.~R.~Hu and B.~M\"uller,
  Phys.\ Lett.\  B {\bf 409}, 377 (1997);
G.~D.~Moore, C.~Hu and B.~M\"uller,
  Phys.\ Rev.\  D {\bf 58}, 045001 (1998).

\bibitem{Danielewicz:1991dh}
P.~Danielewicz and G.~F.~Bertsch,
  Nucl.\ Phys.\ \textbf{A533}, 712 (1991);
A.~Lang, H.~Babovsky, W.~Cassing, U.~Mosel, H.~Reusch, and
  K.~Weber, J.\ Comp.\ Phys.\ \textbf{106}, 391 (1993);
Z.~Xu and C.~Greiner, Phys.\ Rev.\ \textbf{C71}, 064901 (2005).

\bibitem{RHICqhat}
A.~Majumder,
  J.\ Phys.\ G {\bf 34}, S377 (2007).

\bibitem{Schenke:2006xu}
B.~Schenke, M.~Strickland, C.~Greiner, and M.~H.~Thoma,
  Phys.\ Rev.\ \textbf{D73}, 125004 (2006).

\bibitem{Armesto:2004vz}
  N.~Armesto, C.~A.~Salgado and U.~A.~Wiedemann,
  Phys.\ Rev.\  C {\bf 72}, 064910 (2005).

\bibitem{DGMV}
A.~Dumitru, F.~Gelis, L.~McLerran and R.~Venugopalan,
  arXiv:0804.3858 [hep-ph].

\end{thebibliography}

\end{document}